**Spin mediated magneto-electro-thermal transport behavior in $Ni_{80}Fe_{20}$/MgO/p-Si thin films**

P. C. Lou[1], W. P. Beyermann[2] and S. Kumar[1,3,*]

[1] Department of mechanical engineering, University of California, Riverside, CA

[2] Department of physics and astronomy, University of California, Riverside, CA

[3] Material science and engineering program, University of California, Riverside, CA

**Abstract**

In Si, spin-phonon interaction is the primary spin relaxation mechanism. At low temperatures, the absence of spin-phonon relaxation will lead to enhanced spin accumulation. Spin accumulation may change the electro-thermal transport within the material, and thus may serve as an investigative tool for characterizing spin-mediated behavior. Here we present the first experimental proof of spin accumulation induced electro-thermal transport behavior in a Pd (1 nm)/Ni$_{80}$Fe$_{20}$ (25 nm)/MgO (1 nm)/p-Si (2 μm) specimen. The spin accumulation originates from the spin-Hall effect. The spin accumulation changes the phononic thermal transport in p-Si causing the observed magneto-electro-thermal transport behavior. We also observe the inverted switching behavior in magnetoresistance measurement at low temperatures in contrast to magnetic characterization, which is attributed to the canted spin states in p-Si due to spin accumulation. The spin accumulation is elucidated by current dependent anomalous Hall resistance measurement, which shows a decrease as the electric current is increased. This result may open a new paradigm in the field of spin-mediated transport behavior in semiconductor and semiconductor spintronics.

**Introduction**

In silicon, electron-phonon scattering via the Elliott-Yafet mechanism is believed to be the primary spin relaxation mechanism at room temperature[1,2]. At low temperatures, phonon occupation is significantly reduced, which changes the spin relaxation behavior and increases the spin diffusion length. We hypothesized that at low temperatures the reduction in spin-phonon relaxation will lead to non-equilibrium spin accumulation in the bulk in addition to that at the surfaces/interfaces. With this large spin accumulation, spin-phonon interactions may change the electro-thermal transport behavior in non-magnetic materials (Si in the present study), especially when the spin-orbit coupling is small since the large spin-orbit coupling will lead to dephasing. Such a change in behavior can be elucidated by a change in phonon-mediated thermal properties.

**Experimental details**

For this paper, we used the self-heating 3ω method[3] to measure in-plane thermal conductivity of thin-film specimens. A schematic of the experimental setup is shown in Figure 1 a. To fabricate the thin-film structure, we started with a commercially available silicon on insulator (SOI) wafer with a B-doped 2 μm thick device layer having a resistivity of 0.001-0.005 Ω cm. Using UV photolithography and deep reactive ion etching (DRIE), we etched the handle layer underneath the specimen region. Then we patterned and etched the front setup in the Si device layer using DRIE. A freestanding Si structure was made using hydrofluoric acid vapor etching. In the next step, we removed the surface oxide by Ar milling for 10 minutes, followed by adding a 1 nm layer of MgO using RF sputtering. A layer of 25 nm $Ni_{80}Fe_{20}$/1 nm Pd was then deposited onto the device using e-beam evaporation. The material deposition using evaporation leads to line-of-sight thin film deposition on the top of p-Si layer only. The MgO layer acts as a tunneling barrier for spin transport across the interface between the Si and the $Ni_{80}Fe_{20}$/Pd layers. The freestanding specimen with dimensions 170 μm × 9 μm × 2 μm is a multilayer consisting of Pd (1 nm)/ $Ni_{80}Fe_{20}$ (25 nm)/MgO (1 nm)/p-Si (2 μm) as shown in Figure 1 b. Being a metal, $Ni_{80}Fe_{20}$ is significantly more electrically conductive than the p-Si layer, which has a resistivity of 0.001-0.005 Ω cm. We estimated the resistance of the $Ni_{80}Fe_{20}$ layer to be ~290 Ω, while the resistance of the p-Si layer is estimated to be ~390 Ω, which ensures that the current will not be shunted across the metal layer. It needs to be stressed that the 3ω method[4] can be used for insulating thin films and shunting across the metal (heater) layer does not effect the thermal transport measurement. The observation of inverse spin Hall effect (ISHE) in p-Si[5] implies the existence of the spin-Hall effect (SHE) due to reciprocity. The SHE in p-Si[6,7] should lead to polarization of the p-Si layer since spin current is absorbed at the interface, a behavior inferred from the spin-Hall magnetoresistance (SMR)[8,9]. At temperatures above 50 K the Elliot- Yafet spin-phonon interactions are the dominant spin relaxation mechanism as recently proposed by Lou et al.[6]. But at low temperatures, the decreased phonon occupation allows the polarized spin current to develop into a non-equilibrium spin accumulation as shown in figure 1 c. Magneto-electro-thermal transport measurements as a function of temperature presented in this work support the proposed hypothesis.

(Figure 1)

We carried out the magneto-electro-thermal transport measurements with a Quantum Design physical property measurement system (PPMS) using the 3ω method[3] with the sample connected in a 4-probe configuration. The 3ω method utilizes a time-dependent current of frequency ω and amplitude $I_0$ in the specimen to both generate the temperature fluctuations and probe the thermal response. The technique relies on the solution of the one-dimensional heat conduction equation for the specimen, which is given by

$$\rho C_p \frac{\partial \theta(x,t)}{\partial t} = \kappa \frac{\partial^2 \theta(x,t)}{\partial x^2} + \frac{I_o^2 \sin^2 \omega t}{LS}(R_o + R'\theta(x,t)), \tag{1}$$

where $L$ and $S$ are the length between the voltage contacts and the cross-sectional area of the specimen, respectively. $\rho$, $Cp$ and $\kappa$ are the density, specific heat and thermal conductivity in the material. $R_0$ is the initial electrical resistance of the specimen at temperature $T_o$. $R'$ is the temperature derivative of the resistance $R' = \left(\frac{dR}{dT}\right)_{T_o}$ at $T_o$. $\theta(x,t) = T(x,t) - T_o$ is the temporal ($t$) and spatial ($x$) dependent temperature change, as measured along the length of the specimen, which coincides with the heat flow direction. The thermal conductivity can be expressed in terms of the third harmonic voltage $V_{3\omega}$ in the low frequency limit by

$$\kappa \approx \frac{4 I_0^3 R_o R' L}{\pi^4 V_{3\omega} S} \tag{2}$$

We measured the relationship between the applied current (0.9 mA - 1.4 mA) and $V_{3\omega}$ across the specimen at 5 K as shown in Supplementary Figure S1. When the data are fit with power law, the coefficient is 2.78, which is slightly less than 3, which can be attributed to the multilayer structure and thickness of the specimen in this study. In addition, the significantly large magnitude of 1ω compared to 3ω can complicate accurate measurements of the power law relationship. The major sources of error in the thermal conductivity calculation include the dimensional inaccuracies and curve fitting error in the temperature coefficient of resistance. The $R'$ measurement is shown in the Supplementary Figure S2.

**Results and discussion**

The measured resistance and thermal conductivity as a function of the out of plane (z-direction) applied magnetic field at 30 K, 20 K, 10 K and 5 K are shown in Figures 2 a and 2 b.

At these temperatures, phonon occupation is significantly reduced. Contrary to the magnetoresistance, which is temperature independent, the field-dependence of the thermal conductivity changes with temperature. At 30 K, the thermal conductivity increases by ~7.5% for an applied out-of-plane magnetic field of 3 T, and this increases as the temperature is lowered to 5 K where $\Delta\kappa/\kappa$ ~15.4% for the same field. While the ~40% of the charge transport occurs through the p-Si layer but the magnetoresistance behavior of the composite specimen is dominated by the contribution from the $Ni_{80}Fe_{20}$ because the magnetoresistance of p-Si is negligible. Conversely, $Ni_{80}Fe_{20}$ has a thermal conductivity of ~20 W/m·K (approximately for ~25 nm thick[10]), which is significantly lower than that of Si (~80 W/m·K[11]). Hence the observed thermal transport is attributed primarily to the Si layer only. For in-plane conduction, we estimated that the 2 μm p-Si is 6000 times $\left(R_{thermal} = \frac{l}{\kappa A}\right)$ more thermally conducting than the 25 nm $Ni_{80}Fe_{20}$ layer. The thermal transport in Si is phononic[11,12] and we attribute the change in thermal conductivity as a function of magnetic field to spin-phonon interactions in the p-Si specimen.

(Figure 2)

To discover the origin of this magnetic field-dependent electro-thermal transport behavior, we carried out temperature-dependent $R_{1\omega}$, $V_{2\omega}$ and $V_{3\omega}$ measurements at a cooling/heating rate of 0.3 K/min. The $V_{2\omega}$ is related to spin mediated thermoelectric effects including spin-Seebeck effect (SSE)[13,14] and anomalous Nernst effect (ANE)[14]. Figure 3 a shows the resistance as a function of temperature with zero applied magnetic field. The measurement was performed with the following temperature cycle: Cooled from 350 K to 200 K, heated from 200 K to 300 K, cooled from 300 K to 5 K, and finally heated from 5 K to 300 K. As stated earlier, the resistance of the specimen has contributions from both the $Ni_{80}Fe_{20}$ and p-Si layers. We observed the metallic temperature dependence; however, the data displayed hysteretic behavior. We cooled the specimen again from 300 K to 100 K and recorded the resistance for 120 mins to verify the hysteretic behavior as shown in Supplementary Figure S3. From this measurement, we can conclude that instrument temperature drift is the underlying cause of hysteresis in the measurement. While the cooling/heating rate is 0.3 K/min but the specimen is freestanding and device structure is isolated by a 1 µm thick oxide layer, which leads to observed thermal drift. To uncover the temperature dependent magneto-electro-thermal transport behavior, we measured the $V_{2\omega}$ and $V_{3\omega}$ responses as a function of temperature as shown in Figures 3 b-d. We cooled the specimen from around room temperature (350 for $V_{2\omega}$ and 300 K for $V_{3\omega}$) to 5 K at 0.3 K/min. We next applied an out-of-plane magnetic field of 1.25 T, which corresponds to the saturation magnetization of $Ni_{80}Fe_{20}$. Then, we heated the specimen from 5 K to 150 K. From 150 K, we cooled the specimen again to 20 K followed by heating it to 300 K. We observed a sign change in the $V_{2\omega}$ data at ~250 K, followed by a valley at ~220 K. Similarly, we observe a plateau in the $V_{3\omega}$ measurement around ~220 K. We propose that a spin current across the p-Si and $Ni_{80}Fe_{20}$ interface is the origin of the observed $V_{2\omega}$ response since it is a linear function of applied current as opposed to quadratic for SSE and ANE. The linear $V_{2\omega}$ response may occur due to shunting across the bulk of p-Si and $Ni_{80}Fe_{20}$ layers. The sign change in the $V_{2\omega}$ measurement may be attributed to an interfacial phenomenon leading to a change in spin transport across the interface. The plateau observed in the $V_{3\omega}$ measurement corresponds to the temperature where $V_{2\omega}$ shows a valley.

(Figure 3)

We then repeated the temperature-dependent $V_{3\omega}$ measurements with 0 T and 14 T

applied magnetic fields, and these data are shown in Figure 3 d. For these measurements, we cooled the specimen from 300 K to 5 K at 0.3 K/min. We observe that the $V_{3\omega}$ response at 14 T is similar to the zero-field behavior at high temperatures whereas at low temperature $V_{3\omega}$ response is significantly lower than the zero-field behavior. The thermal mass of the 2 μm thick p-Si layer is significantly larger (6000 times) than the other layers in the structure. Hence, the steady state thermal transport measurements are dominated by the bulk behavior of the p-Si layer. The spin accumulation will lead to the spin-phonon interactions in p-Si layer and reduction in thermal conductivity. One of the contradictions in this study is that the spin dephasing should occur at much magnetic field as compared to 14 T. We do not observe a saturation in case of magneto-thermal transport as shown in Figure 2 b. Song and Dery[1] proposed that exchange interactions may exists in donor concentrations of $10^{19}$ cm$^{-3}$ and below 20 K. Such a behavior may exist in the acceptor concentrations of $10^{20}$ cm$^{-3}$ (this study) and below 30 K. This observation leads us to propose that the spin accumulation creates coherent spin states due to weak exchange interactions and reduction in thermal conductivity due to spin polarization. The external applied magnetic field will lead to dephasing and enhancement in thermal transport behavior as observed in this temperature dependent magneto-thermal transport measurement.

To uncover the proximity effect due to the $Ni_{80}Fe_{20}$ layer on p-Si, we carried out temperature dependent magnetization measurements of the Pd (1 nm)/ $Ni_{80}Fe_{20}$ (25 nm)/MgO (1 nm)/p-Si (2 μm) structure using a Quantum Design magnetic property measurement system (MPMS). The magnetic moment was measured while cooling the sample from 300 K to 5 K and then heating it back up from 5 K to 300 K. These data are shown in Figure 4 a. When the applied field is 20 Oe, the measurement shows an anomalous hysteretic behavior between 300 K and 175 K, which happens to coincide with the plateau in the $V_{3\omega}$ measurements. We repeated the temperature dependent magnetic moment characterization with an applied magnetic field of 2000 Oe, which is larger than the saturation magnetic field of the $Ni_{80}Fe_{20}$ thin film. In this case, anomalous behavior disappears. We hypothesized that the thermal hysteretic behavior in magnetic measurement at 20 Oe may originate from the spin mixing at the interface. We measured the magnetic hysteresis as well, and these data are shown in Figure 4 b. A small exchange bias on the order of 10 Oe is seen in the magnetic hysteresis at low temperatures, but this is most likely a consequence of a remnant field in the magnetic solenoid of the instrument. These magnetic measurements demonstrate that the observed behavior originates from transport and not from the proximity effect. The magnetic characterization also precludes any Ni or Fe diffusion into the Si layer.

(Figure 4)

As stated above, we believe the underlying cause of the observed magneto-thermal transport behavior is spin accumulation induced coherent spin states in p-Si. To further elucidate this behavior, we carried out magnetoresistance (MR) measurements using a DC current of 10 µA (~55.5 A/cm$^2$) at multiple temperatures as a function of in-plane transverse magnetic field shown in Figure 4 c and out-of-plane magnetic field shown in Figure 4 d. Increasing the field in the positive direction is indicated with dashed curve, while reversing the direction of the field sweep is indicated with a solid curve. At 300 K (data shown in black with the scale on the left), the sample shows a switching behavior consistent with the anisotropic magnetoresistance (AMR) in $Ni_{80}Fe_{20}$ thin films. But at 5 K (data shown in red with the scale on the right), the switching behavior is anomalous in that the resistance change occurs ~97 Oe before the applied field switches direction instead of after as is the case at 300 K. Similarly, when the field is reduced from the 1500 Oe, switching is observed at ~97 Oe before the applied field switches. This behavior is also observed at 10 K, 20 K and 30 K. The switching behavior is observed for both in-plane transverse and out-of-plane magnetic fields, and it is reminiscent to the inverted hysteresis behavior observed in ferromagnetic/antiferromagnetic composites or multilayers[15]. This behavior may also arise due to spin accumulation induced coherent spin states.

To uncover the origin of spin accumulation, we measured the magneto-electro-thermal transport behavior as a function of electric current as shown in Figure 5 a-b at 20 K. The measurement is carried out at 0.5 mA, 2 mA and 5 mA of heating current. At 5 mA, we estimate the temperature rise to be approximately 80 K. The heating causes the MR to drop to 1.75% as compared to the Figure 2 a. The inverted switching behavior is observed at 2 mA and 5 mA of electric current but not at 0.5 mA. In the current dependent $V_{3\omega}$ response, we observe an inversion at 5 mA as shown in Figure 5 b, which is attributed to the rise in temperature and in turn change in spin-phonon relaxation behavior. We do not observe a change in behavior at 1.25 T (corresponding to saturation magnetization of $Ni_{80}Fe_{20}$) in $V_{3\omega}$ response at 0.5 mA similar to Figure 2 b. But, we observe a change in slope at 1.25 T corresponding to 2 mA and 5 mA heating current. The transport characterization indicates SHE being the mechanism of spin accumulation. In order to support our argument, we fabricated a Hall bar device having Pd (1 nm)/$Ni_{80}Fe_{20}$ (25 nm)/MgO (1 nm)/p-Si (2μm) specimen. We measured the anomalous Hall effect as a function of applied current as shown in Figure 5 c. As the applied electric current is increased, we observe a decrease in $R_{AHE}$ from 11.35 mΩ at 0.5 mA to 9.11 mΩ at 2 mA and 7.07 mΩ at 5 mA. This reduction of ~37.7% cannot occur due to heating since we expect a temperature rise of ~80 K as stated earlier. From the temperature dependent magnetization behavior presented in Figure 4 a, the reduction in magnetic moment from 20 K to 100 K is only ~2.5%. This behavior can only arise due to spin accumulation in p-Si leading to net magnetic moment opposing the magnetic moment of $Ni_{80}Fe_{20}$. The $Ni_{80}Fe_{20}$ layer is essential to observe spin polarization induced magneto-thermal transport behavior. The observed magneto-thermal transport behavior disappears in the absence of ferromagnetic layer[6,7].

We undertake the angular field rotation in yz-plane to elucidate the spin dependent magneto-electro-thermal transport behavior. We carried out the measurement at 100 K and 5 K. We observe AMR behavior in the resistance measurement, which originate from Ni80Fe20 layer (Supplementary Figure S4). The absence of spin-Hall magnetoresistance (SMR) is attributed to the small spin-Hall angle ($10^{-4}$) of p-Si. In addition, we observe a $\sin(\phi_{zy})$ behavior in the $V_{2\omega}$ response measured at 1 T, 4 T and 8 T. This behavior can arise from SSE or ANE. The amplitude of the sine function shows an increase with applied magnetic field, which is not expected for SSE. In addition, the small spin Hall angle, as stated earlier, of p-Si precludes the existence of SSE. We propose that this behavior arises from ANE. The observation of ANE signifies the existence of vertical temperature gradient. This vertical temperature gradient may change the direction as the temperature is reduced leading to the change in the sign of $V_{2\omega}$ response observed in temperature dependent measurement. We also observe a large background $V_{2\omega}$ response in these measurements, which may arise due to thermal fluctuations due to spin polarization and spin-phonon interactions.

We have established that the spin accumulation induced coherent spin state in p-Si is the underlying cause of the observed behavior. We propose that there are two possible spin configurations. In the first case, the spin accumulation due to SHE leads to a ferromagnetic phase near the interface, which is antiferromagnetically coupled to the moment of $Ni_{80}Fe_{20}$. This behavior is similar to inverted hysteresis loop reported by Maity et al. [15] reported in $Ni_{45}Fe_{55}$ thin films. They attributed it to a bimodal distribution of $Ni_{45}Fe_{55}$ and $Ni_3Fe$ grains and due to exchange bias between $Ni_{45}Fe_{55}$ and $Ni_3Fe$ at the grain boundaries. In this study, the MgO spacer layer will eliminate exchange bias at the interface. In addition, we do not observe inverted switching behavior at high temperatures. This behavior eliminates the exchange bias mediated inverted switching mechanism in this study. In the second mechanism, the spin accumulation causes canted AFM spin states in p-Si having net magnetic moment opposite to $Ni_{80}Fe_{20}$ magnetic moment, especially near the interface. At high field, the coherent spin states in p-Si are aligned with the external magnetic field. We propose that the coherent spin states in p-Si relax to canted states as the applied magnetic field is reduced generating a spin current causing inverted switching behavior observed in MR measurement (Supplementary Figure S5). To prove this hypothesis, we undertook the MR measurement from 1500 Oe -1500 Oe (Supplementary Figure S6) at 1.2 mA of heating current. This measurement shows absence of inverted switching behavior supporting our hypothesis of canted coherent spin states due to spin accumulation. Based on the experimental evidence, we propose that the SHE in proximity to a spin porous interface leads to spin polarization of p-Si layer. The spin polarization leads to coherent canted spin states at low temperatures. The relaxation of canted spin states leads to the inverted switching behavior in MR. This unusual behavior is primarily a transport-mediated phenomenon since magnetic measurements do not show an inverted switching behavior.

**Conclusion**

In conclusion, we report measurements of the electrical and thermal transport and the magnetic properties as a function of temperature and applied magnetic field for Pd (1 nm)/ $Ni_{80}Fe_{20}$ (25 nm)/MgO (1 nm)/p-Si (2 μm) thin films. This behavior is attributed to the spin accumulation mediated AFM like spin states in the p-Si layers. The spin accumulation originates from a spin-Hall effect in the p-Si layer in proximity to the porous spin filter at the interface with Ni$_{80}$Fe$_{20}$. The canted coherent spin states lead to the inverted switching behavior seen in MR measurements. In addition, the canted spin states due to spin-Hall effect leads to reduction in anomalous Hall resistance as the applied heating current is increased.

List of Figures:

Figure 1. a. The schematic of experimental setup for magneto-thermal transport measurement with freestanding specimen, b. the SEM micrograph of the MEMS experimental device and c. the schematic of the proposed hypothesis showing canted spin states due to spin accumulation.

Figure 2 a. the magnetoresistance behavior as a function of temperature (30 K, 20 K, 10 K and 5 K) and b. the change in thermal conductivity as a function of magnetic field at multiple (30 K, 20 K, 10 K and 5 K) temperatures.

Figure 3. a. The temperature dependent resistance $R_{1\omega}$ under zero applied magnetic field. b. The second-harmonic voltage $V_{2\omega}$ response as a function of temperature for zero and 1.25 T. c. The third-harmonic voltage $V_{3\omega}$ response as a function of temperature with zero and applied magnetic fields showing the thermal drift and d. The third-harmonic voltage $V_{3\omega}$ response as a function of temperature for applied magnetic field of 0 T and 14 T.

Figure 4. a. The temperature dependent longitudinal magnetic moment for the $Ni_{80}Fe_{20}$/MgO/p-Si thin films. b. The magnetic hysteresis for thin film at 300 K and 5 K. The dc magnetoresistance at 300 K and 5 K for an applied magnetic field c. transverse in-plane (y-direction) and d. out of plane (z-direction).

Figure 5. a. the magnetoresistance behavior at 20 K as a function of applied electric current, b. the $V_{3\omega}$ response at 20 K as a function of applied electric current, c. Hall resistance at 20 K as a function of electric current showing decrease in $R_{AHE}$ and d. the $V_{2\omega}$ response as a function of angular rotation of magnetic field in the zy-plane at 100 K and 5 K.

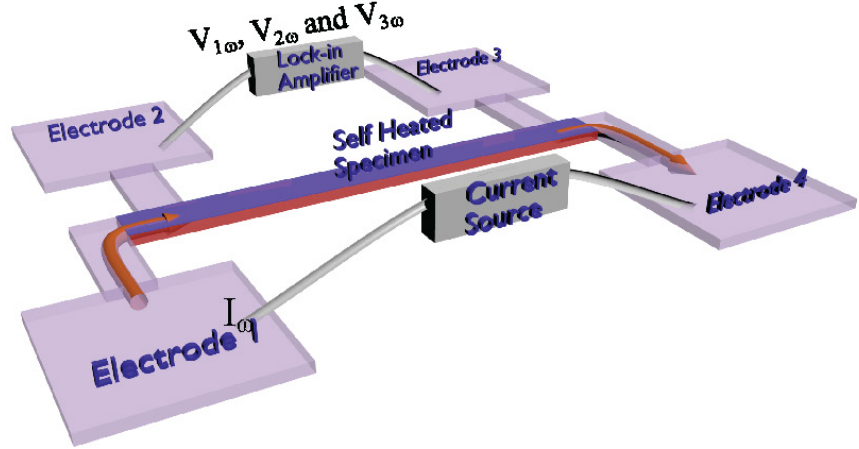

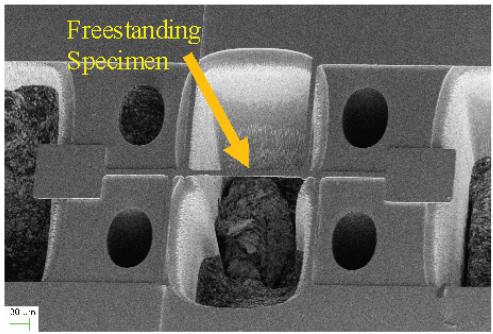

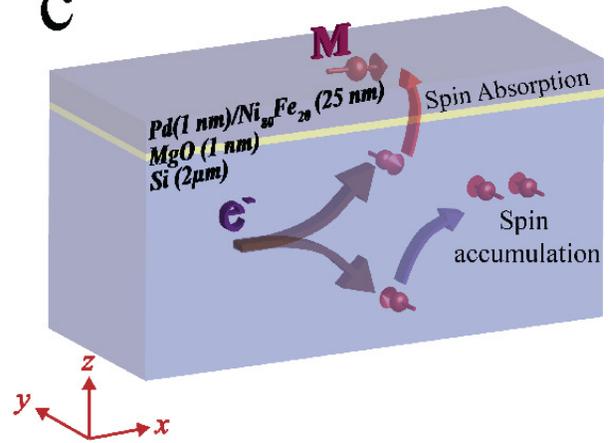

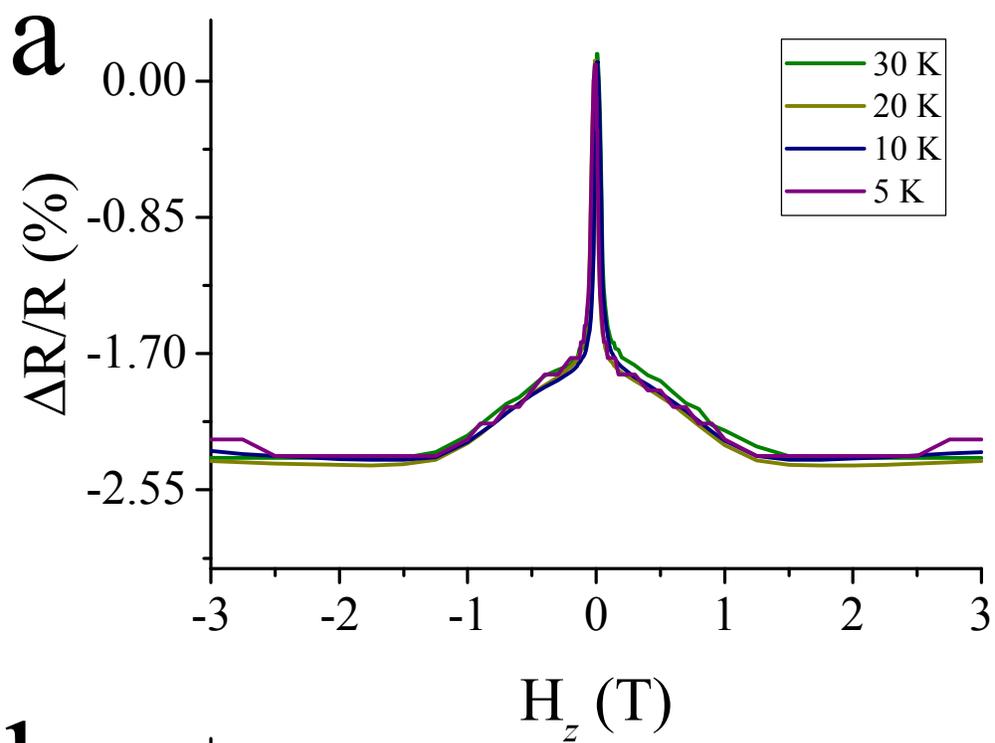

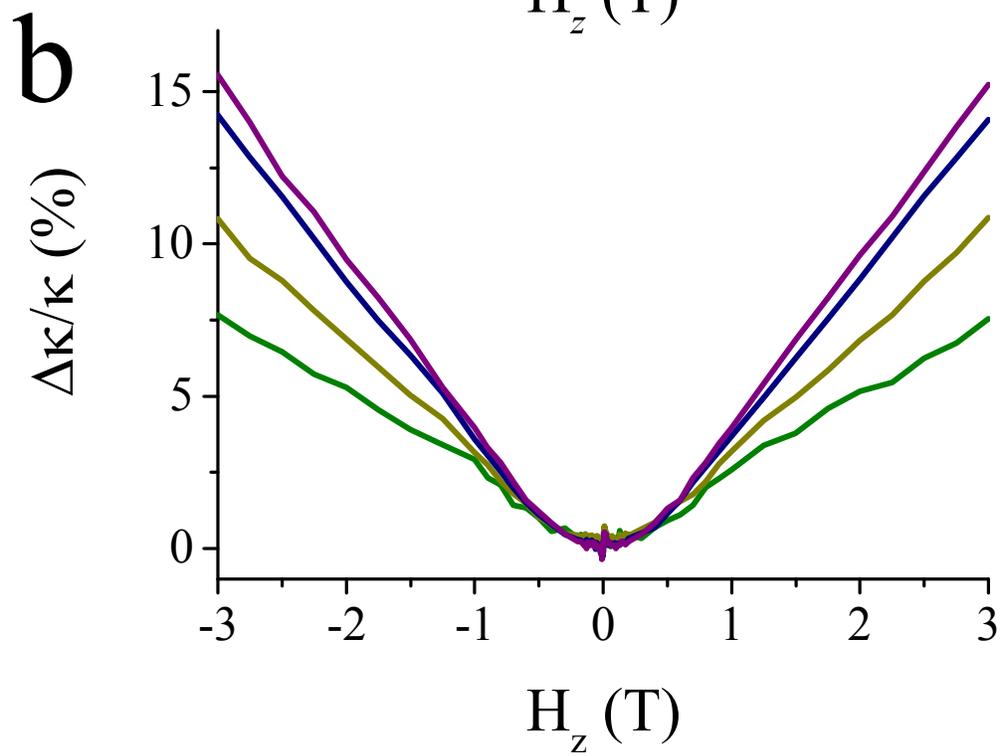

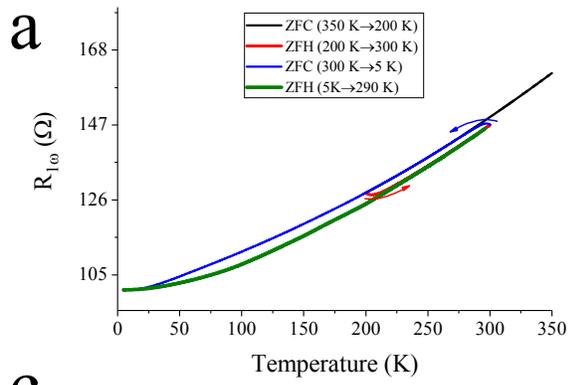
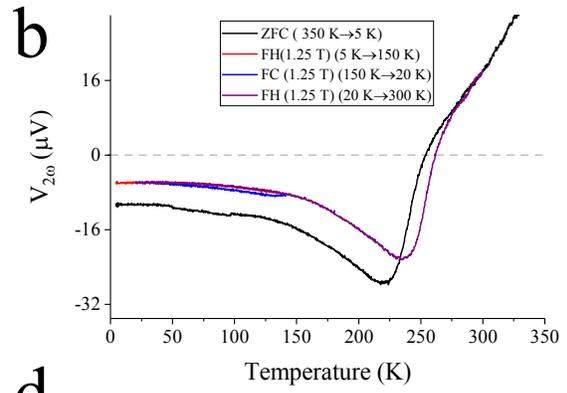
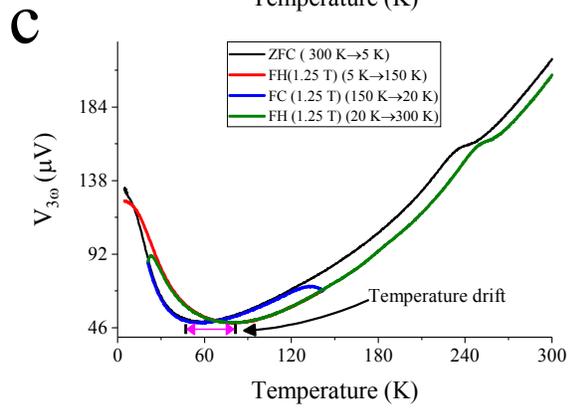
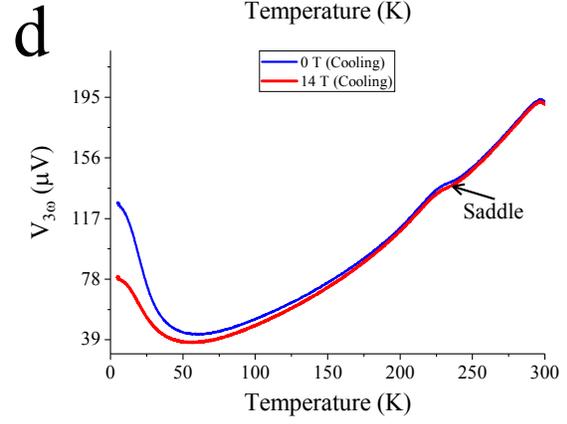

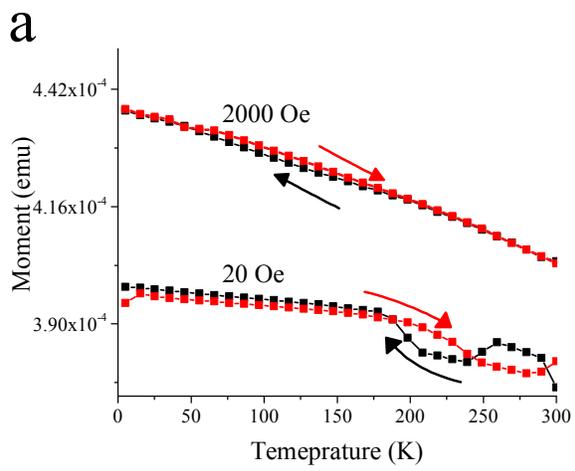
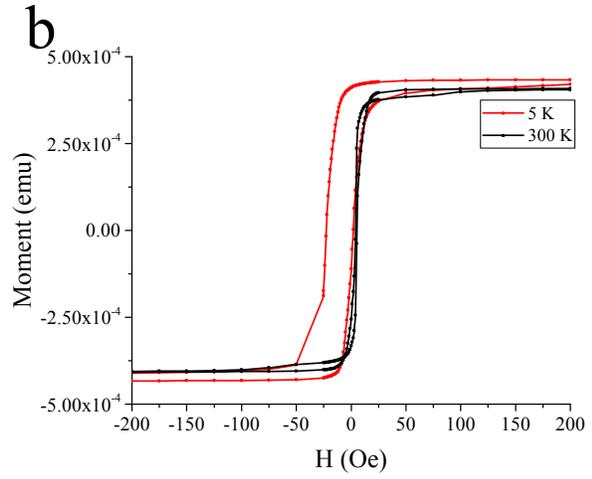
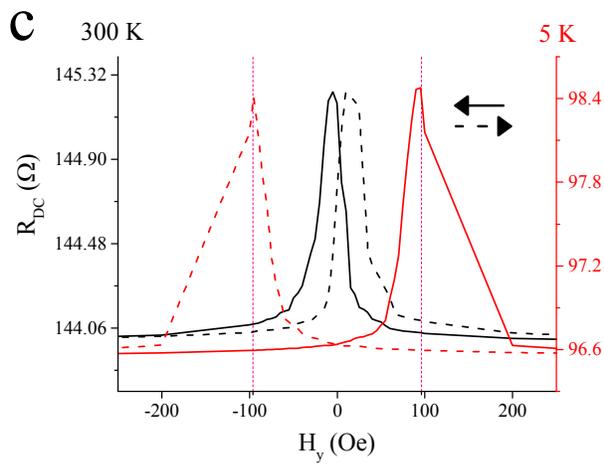
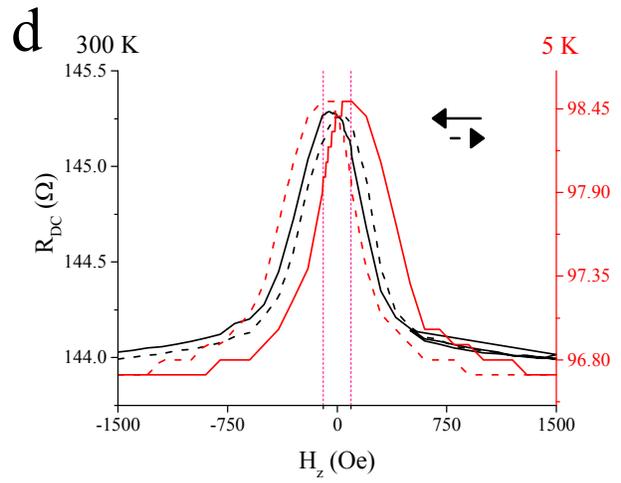

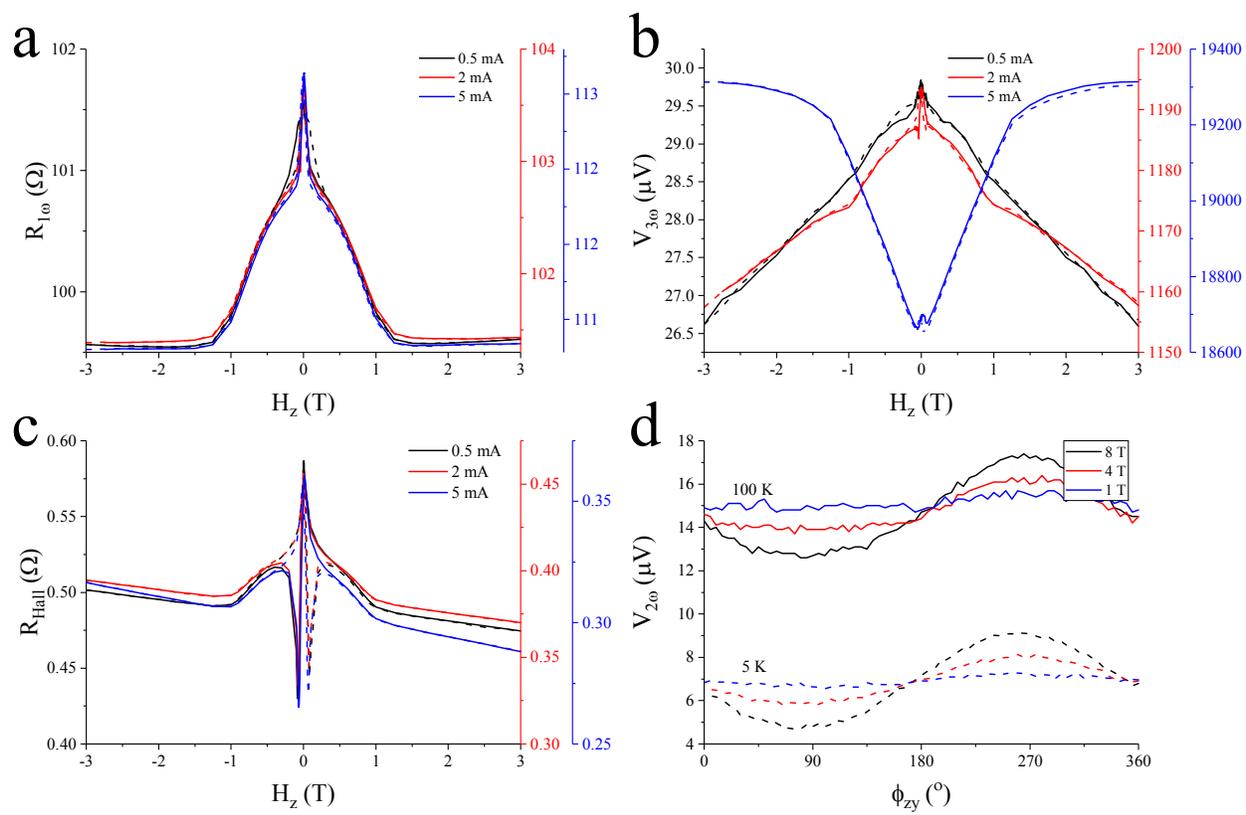

**Supplementary Figures-**

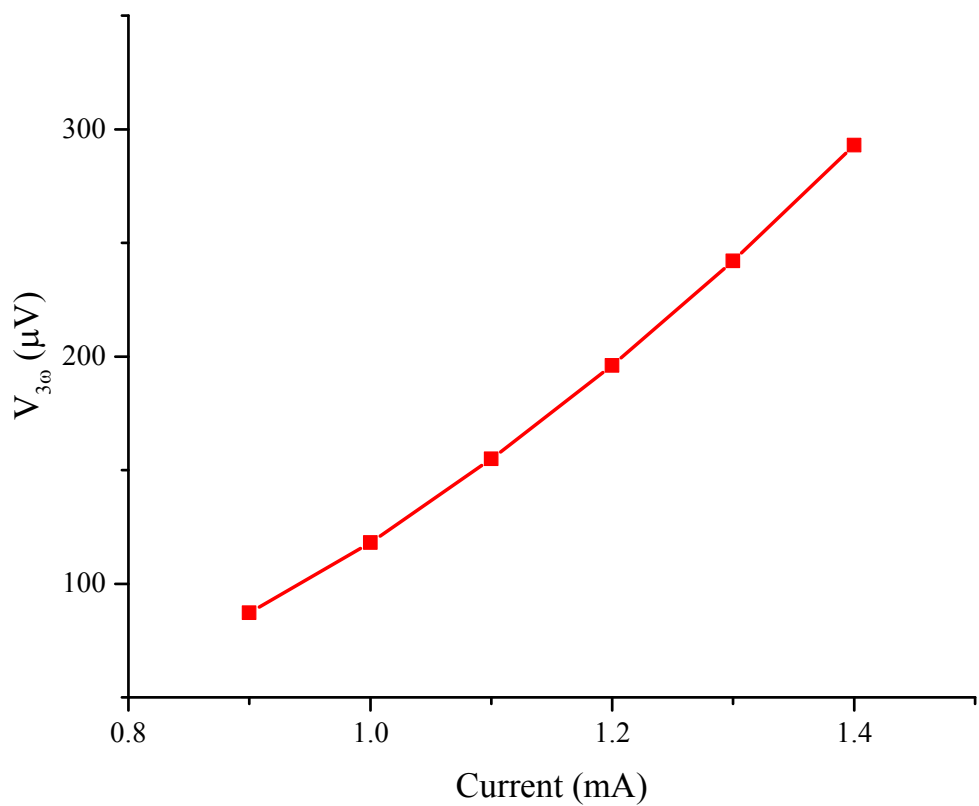

Figure S1. The cubic relationship between current and V$_{3\omega}$ at 5 K.

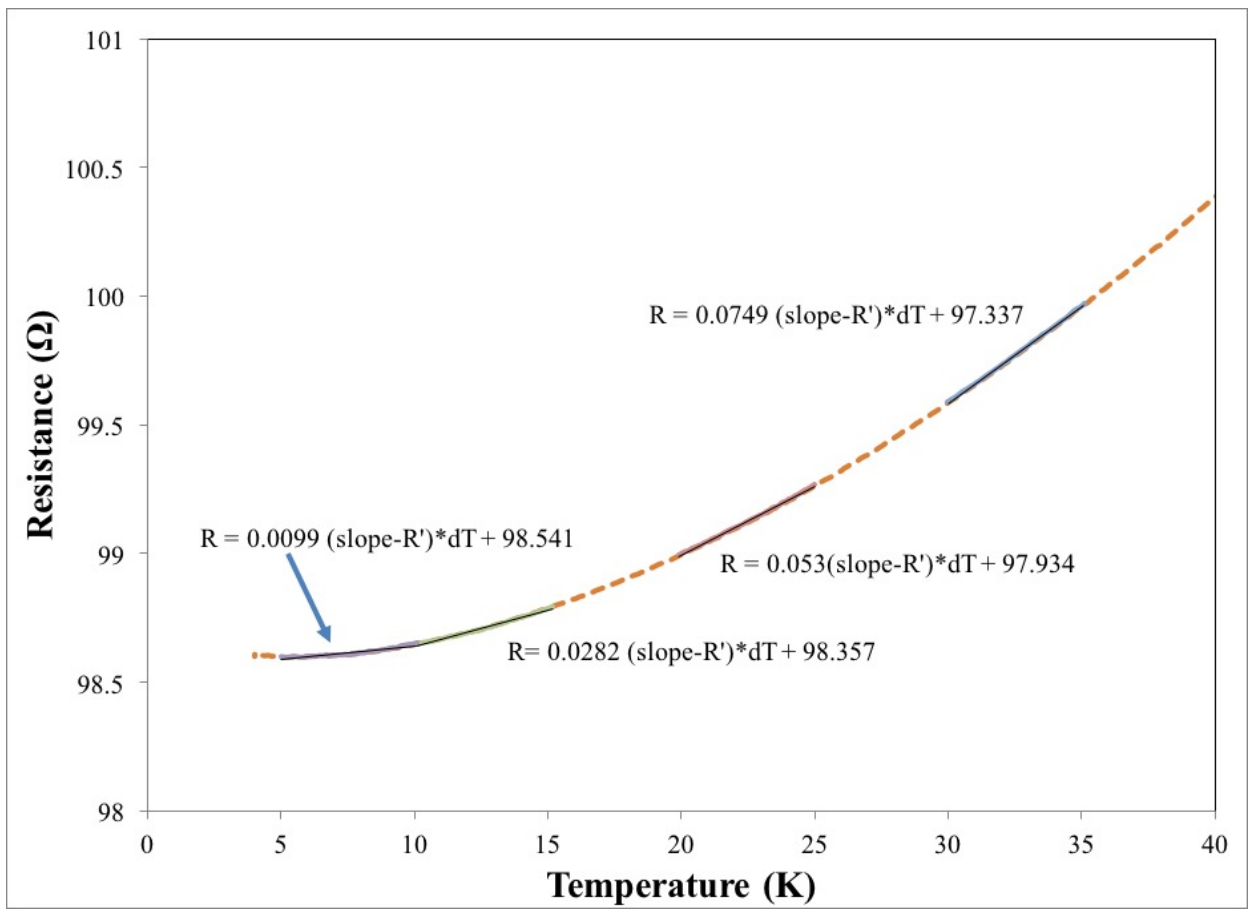

Figure S2. The dR/dT data and four linear fits to the data over temperature ranges starting at 30 K, 20 K, 10 K and 5 K.

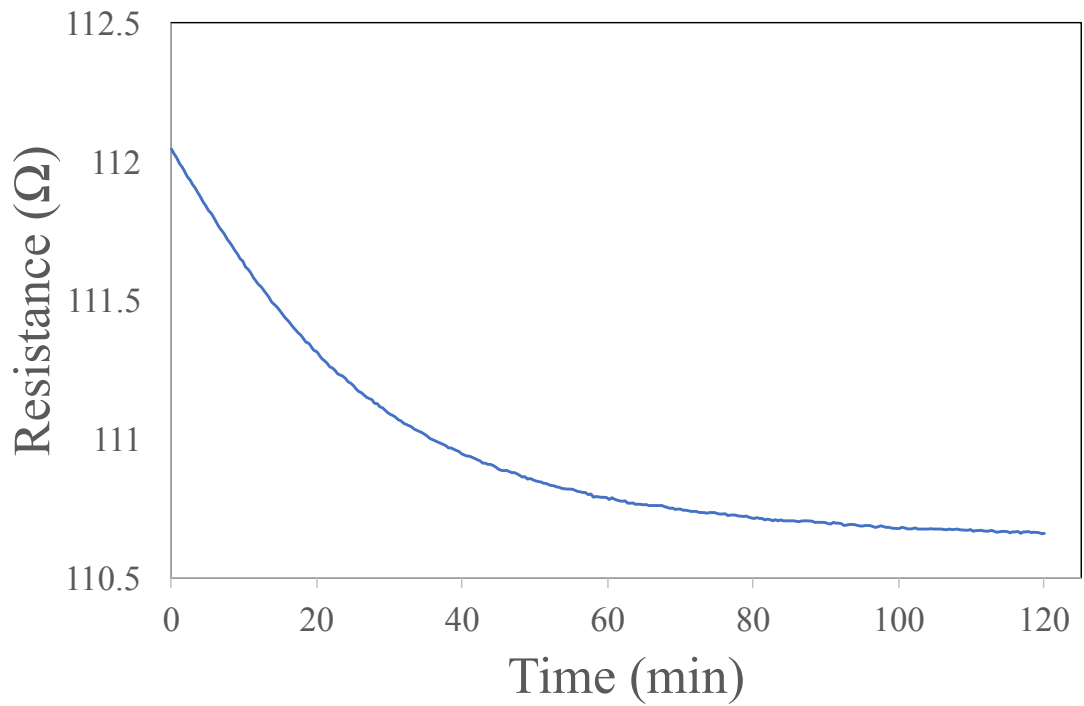

Supplementary Figure S3. The change in resistance due to thermal drift over 120 mins.

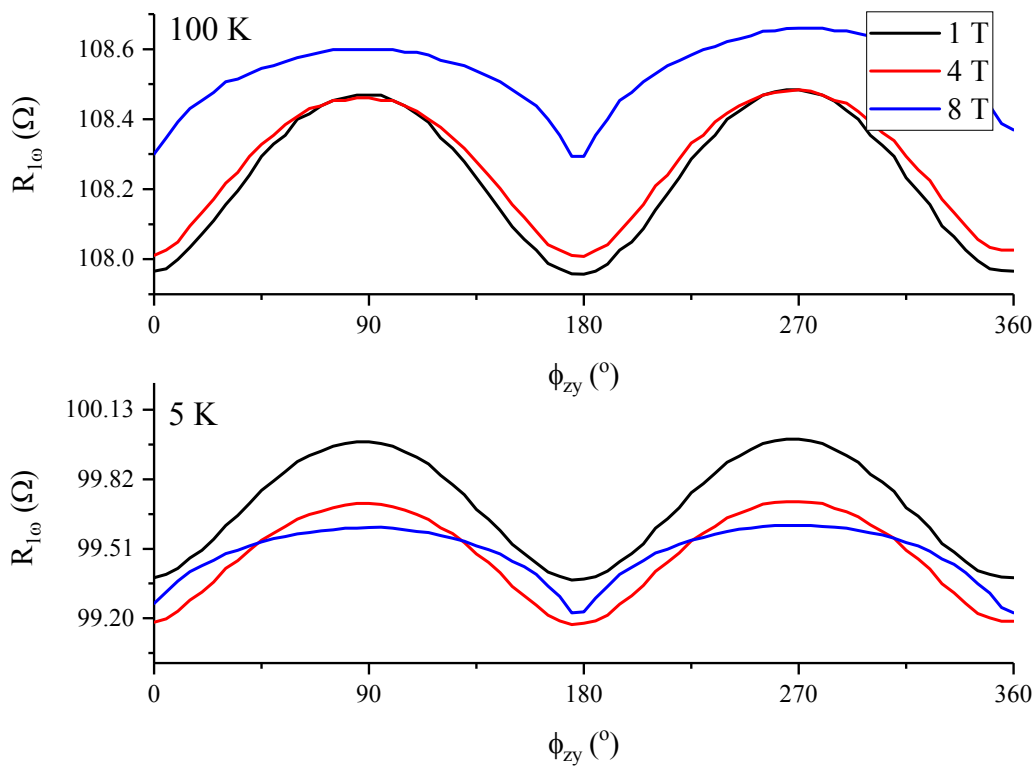

Supplementary Figure S4. The resistance as a function of angular rotation of magnetic field in the yz-plane at 100 K and 5 K.

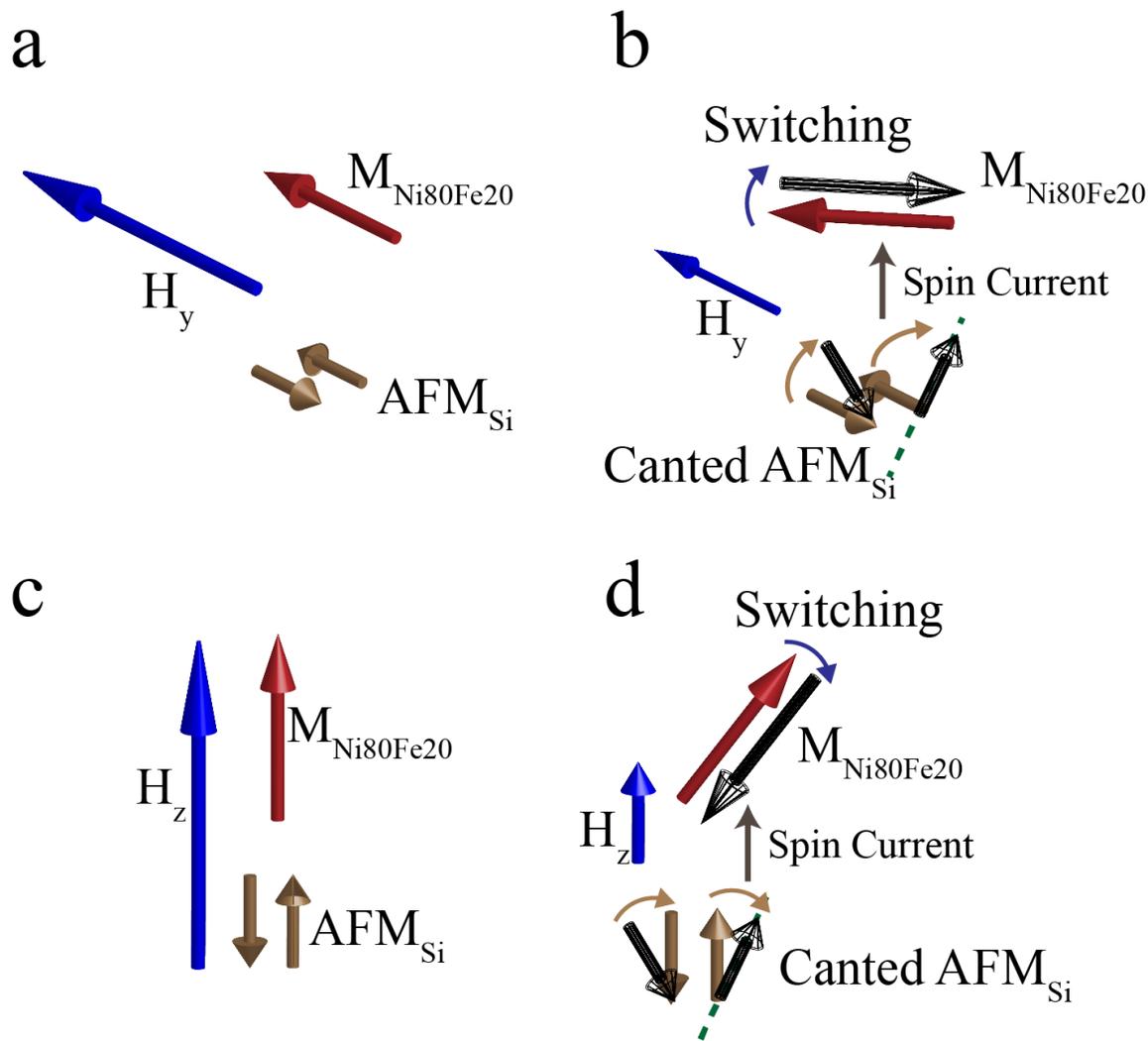

Supplementary Figure S5. The schematic showing the mechanism of inverted switching behavior- for transverse in-plane magnetic field a. at high magnetic field the coherent spin states in Si are aligned with the external magnetic field, b. at low field-coherent spin states relax to canted states generating spin current, which leads to inverted switching behavior. The Similar behavior is proposed for out of plane magnetic field c. at high field and d. low field inverted switching.

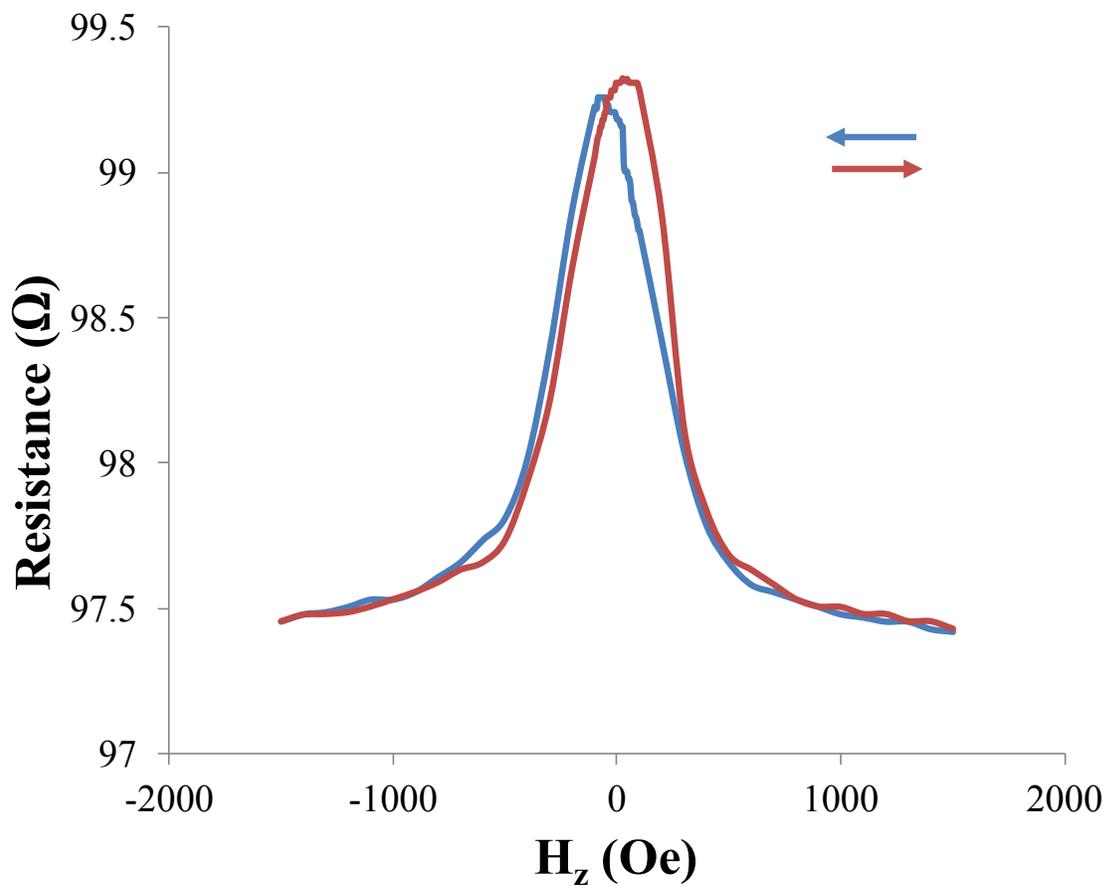

Supplementary Figure S6. The magnetoresistance for applied magnetic field between 1500 Oe and -1500 Oe showing a regular switching behavior. Arrows indicate direction of magnetic field sweep.